%% file: nbody_cosmic_rays_nlt.tex
\documentclass[preprintnumbers,amsmath,amssymb,nofootinbib,twocolumn]{revtex4-1}

\usepackage{dcolumn}
\usepackage{bm}
\usepackage{graphicx}

\usepackage[linktocpage=true,unicode=true]{hyperref}
\usepackage{hypcap}

\usepackage{color}
\usepackage{natbib}
\input{newcommands.tex}

\graphicspath{{Figs/}}

\begin{document}

\title{Indirect dark matter searches: Towards a consistent top-bottom approach for studying 
  the gamma-ray signals and associated backgrounds}

\author{Emmanuel Nezri}
\affiliation{Aix Marseille Université, CNRS, LAM (Laboratoire d'Astrophysique de 
Marseille) UMR 7326, 13388, Marseille, France}
\email{Emmanuel.Nezri@oamp.fr}

\author{Julien Lavalle}
\affiliation{Laboratoire Univers \& Particules de Montpellier (LUPM),
  CNRS-IN2P3 \& Universit\'e Montpellier II (UMR-5299),
  Place Eug\`ene Bataillon,
  F-34095 Montpellier Cedex 05 --- France}
\email{lavalle@in2p3.fr}

\author{Romain Teyssier}
\affiliation{IRFU/SAp, CEA Saclay, F-91191 Gif-sur-Yvette Cedex --- France\\ }
\affiliation{Institute of Theoretical Physics, University of Z\"urich, 
Winterthurerstrasse 190, 8057 Z\"urich --- Switzerland}
\email{romain.teyssier@cea.fr}

\begin{abstract}
While dark matter is the key ingredient for a successful theory of structure formation, its
microscopic nature remains elusive. Indirect detection may provide a powerful test for some 
strongly motivated DM particle models. Nevertheless, astrophysical backgrounds are usually 
expected with amplitudes and spectral features similar to the chased signals. On galactic scales, 
these backgrounds arise from interactions of cosmic rays (CRs) with the interstellar gas, both 
being difficult to infer and model in detail from observations. Moreover, the associated 
predictions unavoidably come with theoretical errors, which are known to be significant. 
We show that a trustworthy guide for such challenging searches can be obtained by exploiting the 
full information contained in cosmological simulations of galaxies, which now include baryonic 
gas dynamics and star formation. We further insert CR production and transport from the identified 
supernova events and fully calculate the CR distribution in a simulated galaxy. We focus on 
diffuse gamma rays, and self-consistently calculate both the astrophysical galactic emission 
and the dark matter signal. We notably show that adiabatic contraction does not necessarily induce 
large signal-to-noise ratios in galactic centers, and could anyway be traced from the astrophysical 
background itself. We finally discuss how all this may be used as a generic diagnostic tool for 
galaxy formation.
\end{abstract}

\maketitle
\preprint{LUPM:12-019}

\begin{figure*}[t]
\centering
\includegraphics[width=0.265\textwidth]{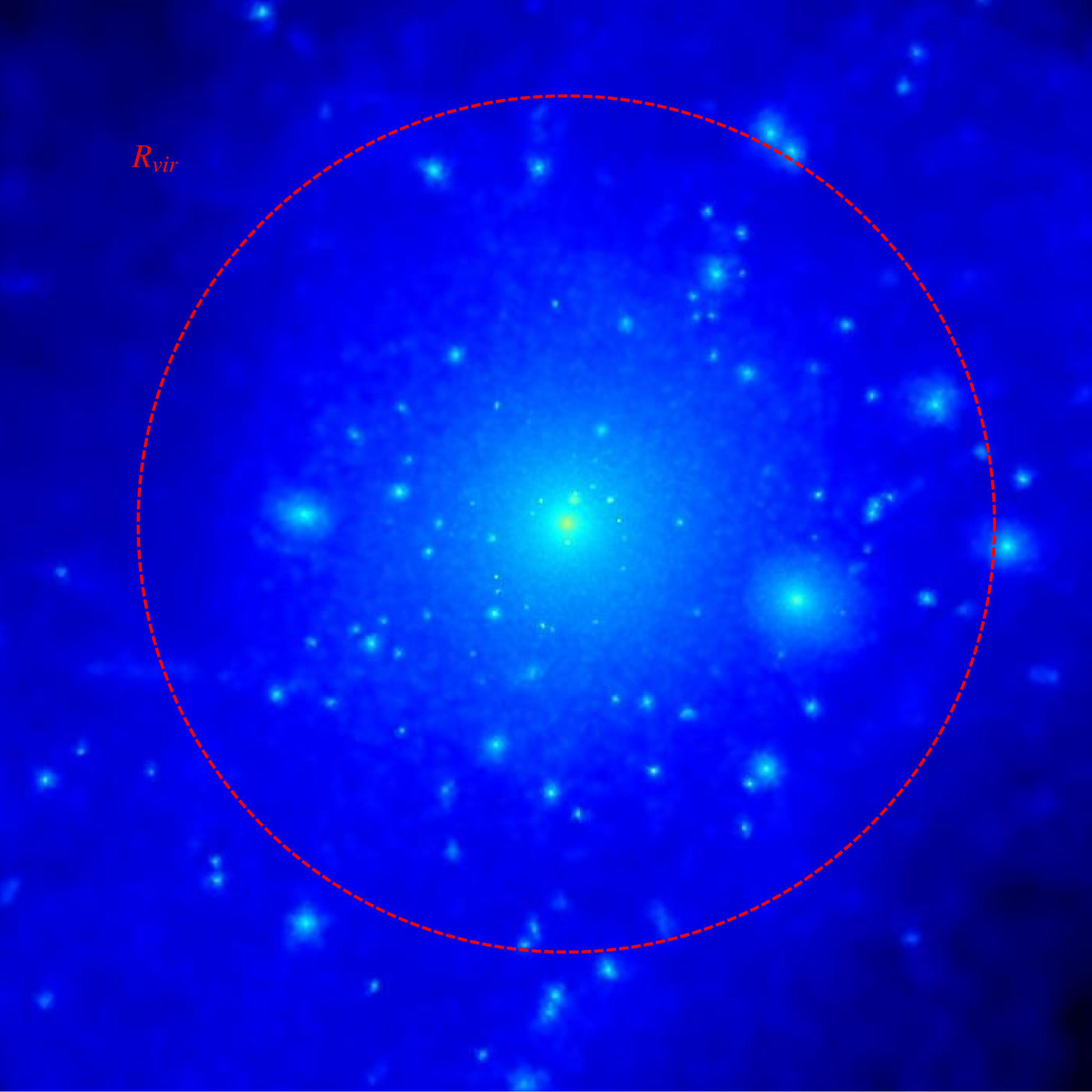}
\includegraphics[width=0.335\textwidth]{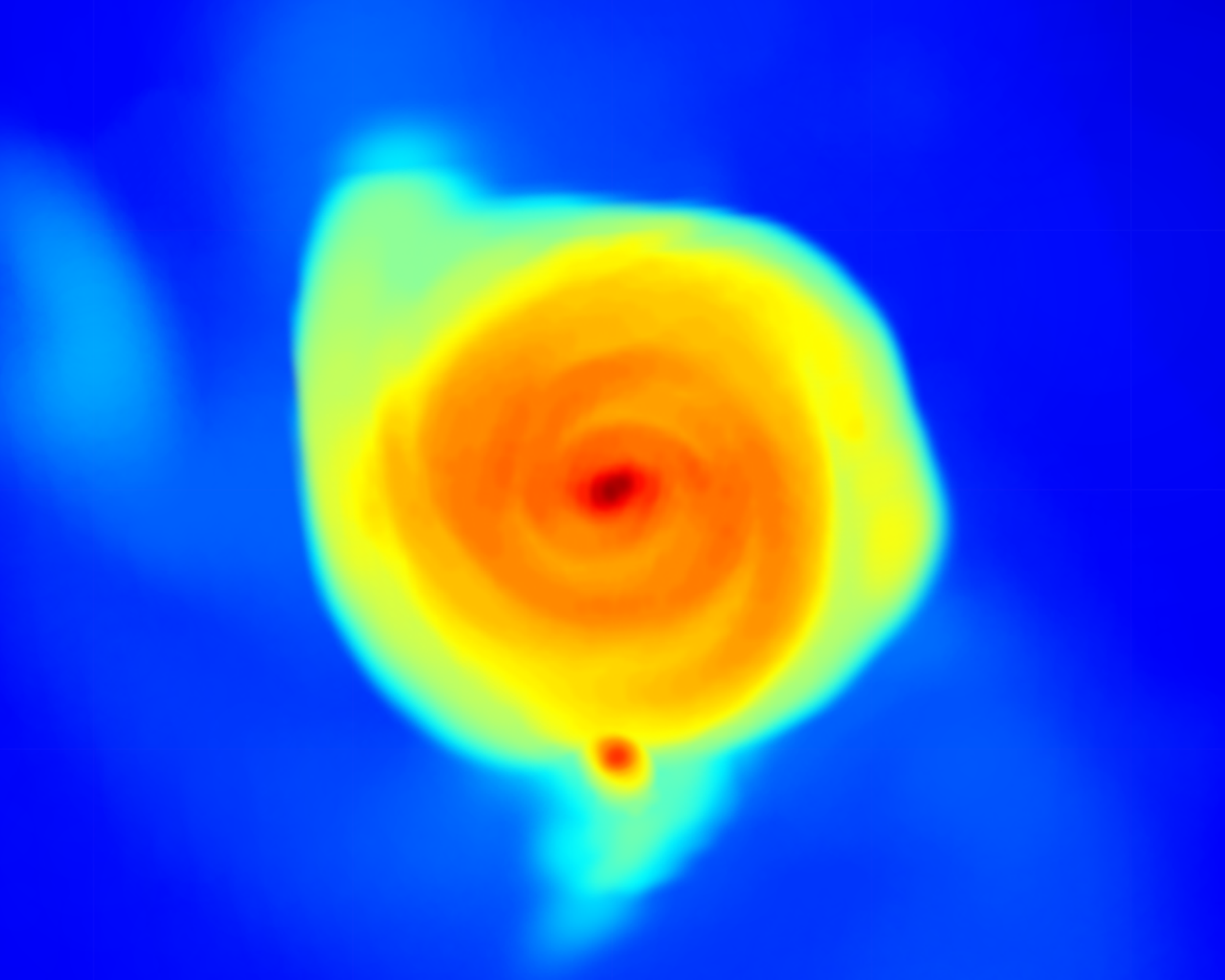}
\includegraphics[width=0.335\textwidth]{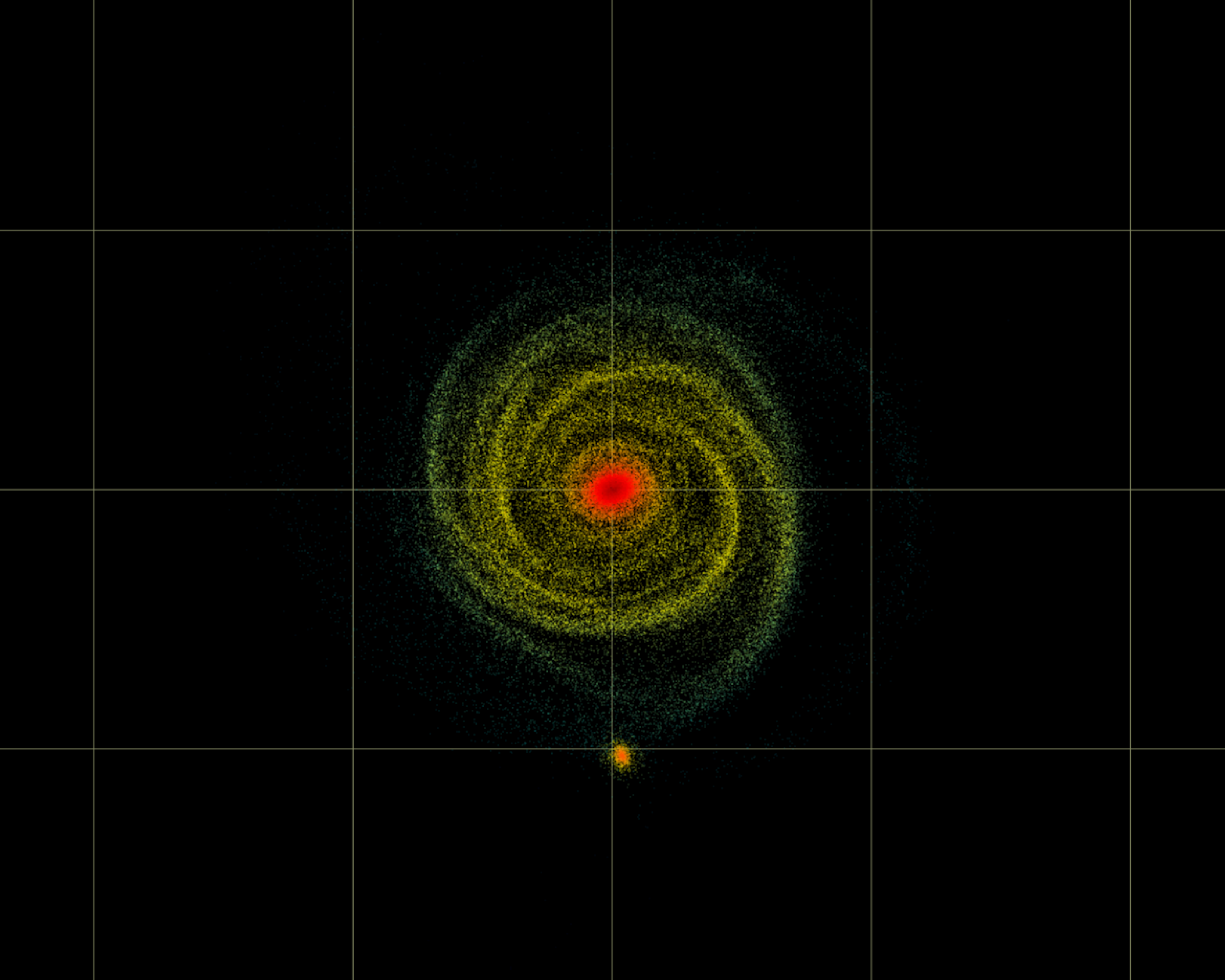}
\caption{Left: DM halo and subhalos; the virial radius (264 kpc) appears as a red circle. Middle: 
top view of the gas content (scaled as in right panel). Right: SN events in the last 500 Myr (10 kpc
grid).}
\label{fig:simu}
\end{figure*}
An important issue arising in indirect dark matter (DM) searches is our limited capability for 
predicting
the astrophysical backgrounds accurately enough (for reviews, see \cite{1996PhR...267..195J,*2000RPPh...63..793B,*2006RPPh...69.2475C,*Salati:2007zz}). In the last decade, several claims for
smoking guns have shown up in the literature, usually followed (sometimes preceded) by 
conventional astrophysical explanations (\ie~other astrophysical sources or wrong background 
models). For examples at the Galactic scale, one may find the {\em WMAP haze}
\cite{2007PhRvD..76h3012H}, which might be due to an improper background extrapolation 
\cite{2010JCAP...10..019M}, or at the cosmic positron excess \cite{2009Natur.458..607A}, which 
triggered a plethora of DM proposals (\eg~\cite{2008PhRvD..78j3520B}), while pulsars have long been 
known to be good candidates (\eg~\cite{1987ICRC....2...92H}).

Gamma rays are probably the best messengers for seeking DM annihilation traces
because of the favorable experimental landscape and because they propagate freely from their 
source to the observer, thereby reducing the theoretical uncertainties in predictions with respect 
to charged species. The best targets are those which can be localized accurately and are not 
background polluted: nearby dwarf galaxies. Nevertheless, current 
experimental sensitivities fall just too short \cite{2011PhRvL.107x1302A,2012ApJ...747..121A}. 
Another possibility is to examine the diffuse gamma-ray emission (DGRE), for instance its 
angular and/or spectral gradients (\eg~\cite{1998APh.....9..137B,2011PhRvD..83b3518P}). However, it 
is often difficult to interpret the data because of uncertainties in Galactic background models (see
\cite{2012ApJ...750....3A} in light of \cite{2011A&A...531A..37D}). One way around this is to focus 
on spectral features specific to DM models, like gamma-ray lines or spectral hardenings 
due \eg\ to the prominence of bremsstrhalung annihilation diagrams~\cite{2005PhRvL..94q1301B,*2008PhRvD..78j3520B,*2012JCAP...07..054B}. Unfortunately, 
current experiments do not yet have the required energy resolution to achieve a good enough 
background rejection, and no such features have been discovered so far. Therefore, a much deeper 
understanding of the astrophysical backgrounds seems now necessary to go from speculation to 
stronger evidence. We focus on gamma rays in the following.

The main astrophysical background comes from interactions of CR
nuclei (mostly protons) with the interstellar gas. It also originates in inverse Compton
scattering of CR electrons off the interstellar radiation fields and the cosmic
microwave background, but we disregard this subdominant component here \new{($\lesssim 10$\% of the 
DGRE measured by Fermi at 1 GeV \cite{2012ApJ...750....3A})}. Hence, a complete prediction
of the DGRE depends on both the Galactic gas and CR distributions. For the former, models are based 
on line-of-sight observations (21 cm for instance) that must be 
deconvolved in agreement with the global Galactic dynamics, a difficult reconstruction procedure 
subject to ambiguities (\eg~\cite{2008ApJ...677..283P}). For the latter, 
one should in principle know about the real distribution of the CR sources in space and time, the 
precise mechanism of acceleration and escape, and the physics of CR transport in the interstellar 
medium. While the latest points still call for theoretical developments, the first one, 
one of the most critical, remains difficult to infer from observations.

A solution to get more insights about these issues is to {\em build} a template galaxy from first 
principles, wherein all ingredients relevant for testing indirect detection are found with
the correct physical correlations. This has actually become possible since the advent of 
cosmological simulations of galaxies including baryons and star formation 
(so-called {\em zoom-in} cosmological simulations\footnote{The {\em zoom-in} technique consists 
of identifying a candidate galactic halo in a low resolution DM-only simulation, and then 
re-simulating it with baryons and with a much better resolution at the candidate halo position.}). 
Indeed, the resulting virtual 
objects are cosmologically and dynamically self-consistent, and while star formation is still 
treated semiempirically, they offer a perfect environment to assess the discovery potential of 
indirect DM detection. Beside fully characterizing the DM and baryonic gas contents, one can also 
trace the star formation history and localize the supernova (SN) events in space and time. From 
these SNRs, a still missing step is to plug in the injection of high-energy CRs and their transport 
further away in the galaxy. A fraction of the SNe energy is actually already used in the form of 
feedback, a mechanism which regulates the adiabatic contraction of DM~\cite{1986ApJ...301...27B,*2004ApJ...616...16G} due to the cooling of baryons---the same mechanism is also responsible 
for CR acceleration. Once the CR distribution is known, the astrophysical DGRE can be calculated and
compared to the DM annihilation signal which scales like the squared DM density.

\begin{figure*}[t]
\centering
\includegraphics[width=0.49\textwidth]{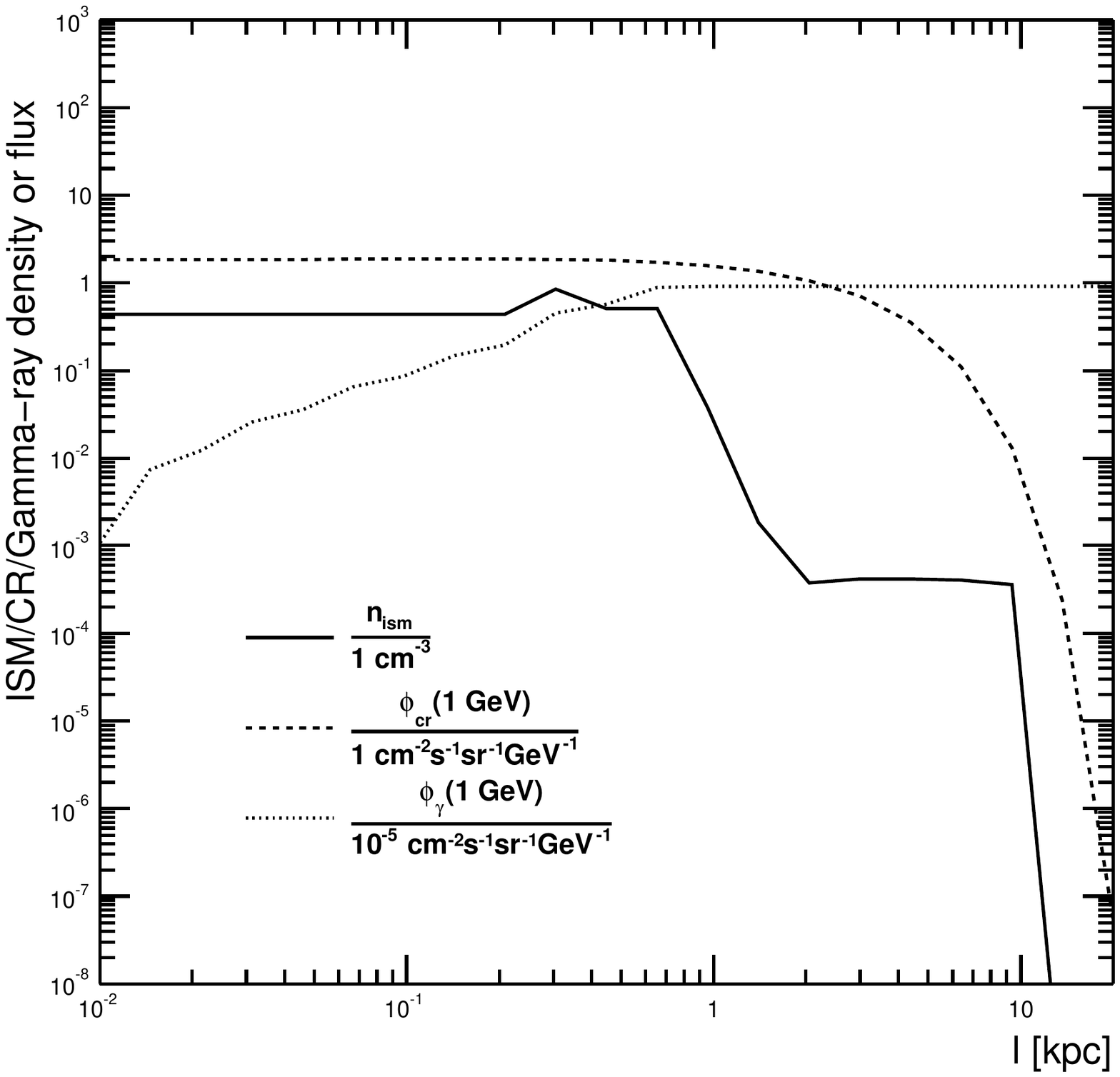}
\includegraphics[width=0.49\textwidth]{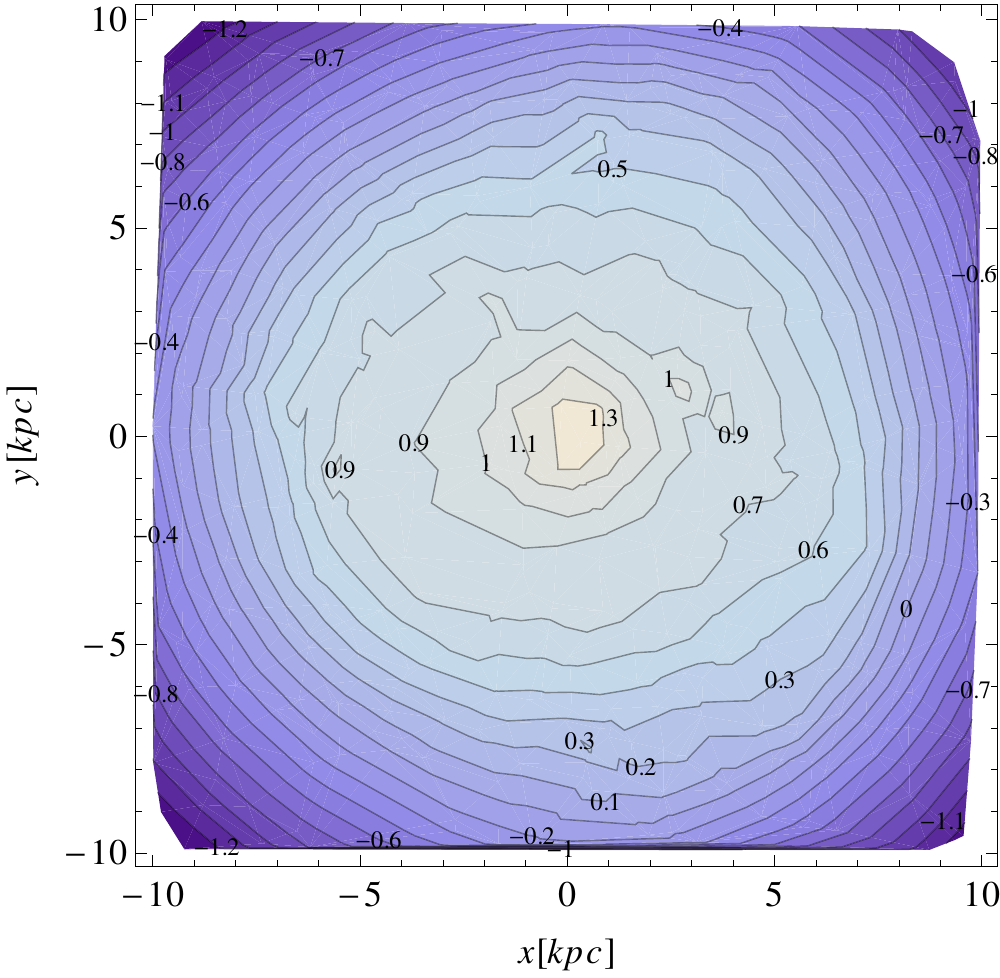}
\caption{Left \new{(units in the legend)}: Vertical gradients of the ISM gas (solid curve) and 
of the CR flux (dashed curve).
The resulting cumulative gamma-ray flux for an observer located at coordinates $(x,y,z)=(0,8,0)$ kpc
is shown as the dotted curve. Right: CR flux gradient in the galactic plane
(in ${\rm log}_{10}\{\Phi_{\rm cr}(>1{\rm GeV})/{\rm cm^{-2}s^{-1}sr^{-1}}\}$).}
\label{fig:crs}
\end{figure*}

We use a zoom-in simulation performed with the cosmological adaptive mesh refinement code 
RAMSES~\cite{2002A&A...385..337T}, which includes the baryon gas dynamics and star formation.
Technical details will be found in \cite{2010JCAP...02..012L}.
At redshift $z=0$, the selected galaxy has a virial radius $R_{200} = 264 $ kpc and features a 
baryon disk of $\sim 10$ kpc radius.
The dark halo mass is $M_h \sim 6\times 10^{11} \msun $, while the bulge and disk masses are both 
$\sim 4\times 10^{10} \msun $, which implies a gas concentration larger than in 
the Galaxy. This induces a contraction of the DM density profile in the center, where the 
logarithmic radial slope is found around $-1.5$, though saturating in the inner 200 pc due to 
resolution limits --- a consequence of adiabatic contraction.
We resolve 79 subhalos, carrying 4.8\% of the total mass. The closest object to 
the galactic center lies at a distance of 12.7 kpc. The different components of our simulated 
galaxy are shown in \citefig{fig:simu}.

Our setup suffers some limitations that we shortly discuss here. They mostly
come from our limited spatial resolution of $\sim 200$ pc. First, the vertical extent of the disk 
is not very well-resolved. Moreover, we cannot study in detail the effects of DM subhalos, but it 
is anyway still hard to incorporate the full baryon treatment in simulations as resolved as in 
\eg~\cite{2008Natur.454..735D,*2008Natur.456...73S}. 
Nevertheless, small subhalos are not expected to accrete ordinary matter efficiently so  
results based on semianalytic studies remain valid (\eg~\cite{2011PhRvD..83b3518P} for 
gamma rays and \cite{2008A&A...479..427L} for antimatter CRs). Finally, we note that
subsequent improvements in the feedback treatment have recently allowed for better control
of the overcooling of baryons at halo centers (\eg~\cite{2010Natur.463..203G}), which could 
have led to a flatter and more extended disk, a less prominent bulge, and a less spiky DM profile. 
Nevertheless, our numerical treatment can still be considered robust given the uncertainties 
affecting the implementation of star formation and feedback in cosmological simulations 
\cite{2012MNRAS.423.1726S}. Besides, these limitations do not qualitatively affect the present 
study which proposes a global phenomenological strategy. Our method, that we detail 
below, can easily be applied to more sophisticated cosmological simulations.

We identified all SN events in the latest 500 Myr of the simulation, for which
we recorded the space-time coordinates. This timescale is taken slightly larger than the residence 
time of CRs in the galaxy, which is set by the diffusion coefficient and the size of the reservoir. 
SNe are defined from the fraction of massive stars (with masses fixed to $10 \, \msun$ here),
which amounts to $\sim$ 10\% of all stars. 
These massive stars are given a fixed lifetime of 20 Myr, after which 
they end as SNe, each converting its core mass into pure energy. 
For each SN, about $10^{51}$ erg is transferred to the surrounding material in the form of kinetic 
energy, 10\% of which is used as feedback energy. As a 
supplementary step, we also convert about 10\% of this kinetic energy into high-energy CRs with 
power-law energy spectra. To our knowledge, this stage is still unprecedented on galactic scales 
(see \cite{2012arXiv1203.1038U} for a complementary approach).
We recorded about 85,000 SN-like events (\ie~$\sim$85,000 star {\em particles} have formed). In 
terms of real SNe (end products of $10 \, \msun$ stars), this translates to $\sim 1.5\times 10^7$ 
events, corresponding to an explosion rate of $\sim$3/100 yr.

To get the full CR distribution through the galaxy, we plug in a full semianalytic computation of 
CR transport by means of Green's functions, using the space-time information associated with 
the sources (SNe). We only consider CR protons and Helium ions, \ie\ the most relevant species 
to calculate the hadronic part of the DGRE, with kinetic energies above a few 
GeV. Such an energy threshold allows one to neglect convection and spallation processes when 
ascribing the (time-dependent) transport equation that CRs do obey 
(for details on CR propagation, see \eg~\cite{berezinsky_book_90}). In this regime,
CRs are mostly subject to spatial diffusion. Diffusion should depend on the galactic magnetic field 
properties, but we lack this piece of information in our simulation. \change{While
seeding the magnetic field is still an open issue, we may still envisage to 
incorporate it semi-consistently in future studies by evolving arbitrary primordial seeds such 
that its typical amplitude on galactic scales at redshift 0 is a posteriori consistent with current 
observations 
\cite{2004ApJ...605L..33H,*2008A&A...482L..13D,*2010A&A...523A..72D,*2012MNRAS.422.2152B}}. 
Instead, we assume a rigidity-dependent, isotropic and homogeneous diffusion coefficient, such as 
in most of models of 
Galactic CRs \cite{1998ApJ...509..212S,*2001ApJ...555..585M,*2008JCAP...10..018E}: $K({\cal R}
\equiv |p|/q) = 0.01 \, {\rm kpc^2 / Myr} \,({\cal R} / 1 \, {\rm GV})^{0.7}$. This leads to 
a maximal range of $\sim 5$ kpc for CRs originating from the oldest sources. The injected CR 
spectrum at SN sources (before transport) is taken to be universal, scaling like $E^{-2}$. The DGRE 
is calculated by convolving the CR flux with the interstellar gas density, 
for which we adopt a fixed relative amount of hydrogen:helium of 9:1. A top view of 
the gas distribution is shown in the middle panel of \citefig{fig:simu}. 
For the hadronic cross sections, we used the semianalytic formul\ae~proposed in 
\cite{2006ApJ...647..692K}, with the nuclear weights of \cite{2007NIMPB.254..187N}.

We illustrate our results by assuming a virtual observer located in the disk at position 
$(0,8,0)$ kpc, in Cartesian coordinates (the disk belongs to the $x$-$y$ plane), as a reference 
to a hypothetical observer on Earth. Yet, this is hard to compare with real Galactic data
since our simulated galaxy does not provide a realistic picture of the Milky Way (MW). Nevertheless,
we can delve into useful qualitative discussions and sketch a roadmap for future and more detailed 
analyses. In the left panel of \citefig{fig:crs}, we plot the vertical gradients of (i) the CR
flux measured at 1 GeV, and (ii) the ISM gas density; we also plot the local cumulative photon flux 
at 1 GeV (the hadronic component), proportional to the line of sight integral of the product of the 
two previous quantities. For the hypothetical observer, this would correspond to a set of data 
associated with Galactic coordinates ($l=0^\circ,b=90^\circ$). We see that the gaseous disk 
extends up to 1 kpc, larger than our Galactic disk at the Sun's position, whose half-width is 
$\sim 100$ pc \cite{2001RvMP...73.1031F}. The CR density has a slightly broader and smooth 
vertical distribution, and experiences an exponential cutoff above a few kpc, which is rather 
consistent with the size of the diffusion zone used in the MW. 
\change{This is not due to any vertical boundary condition, but to the maximal range 
CRs can reach within 500 Myr (sources are confined in the disk).}
Finally, the hadronic part of the local DGRE is obviously found to saturate 
when the gas density shrinks. Our results are actually shown as dimensionless ratios, where the
denominators have been taken with values close to what is measured on Earth. Note that these ratios 
are ${\cal O}(1)$. We show the overall CR flux gradient as 
projected in the disk in the right panel of \citefig{fig:crs}. No cylindrical symmetry appears, as 
is usually assumed in most CR models.

\begin{figure*}[t]
\centering
\includegraphics[width=0.32\textwidth]{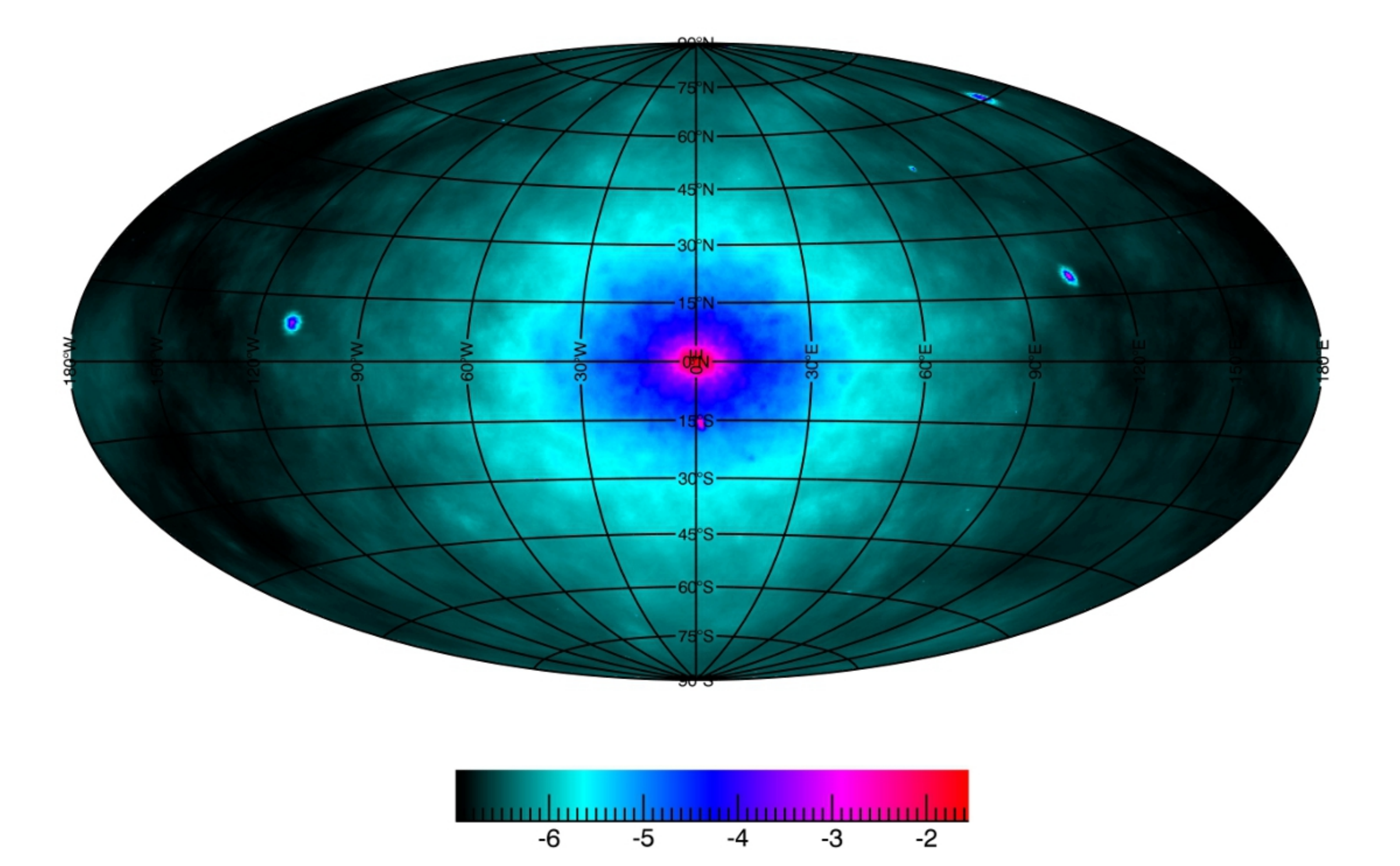}
\includegraphics[width=0.32\textwidth]{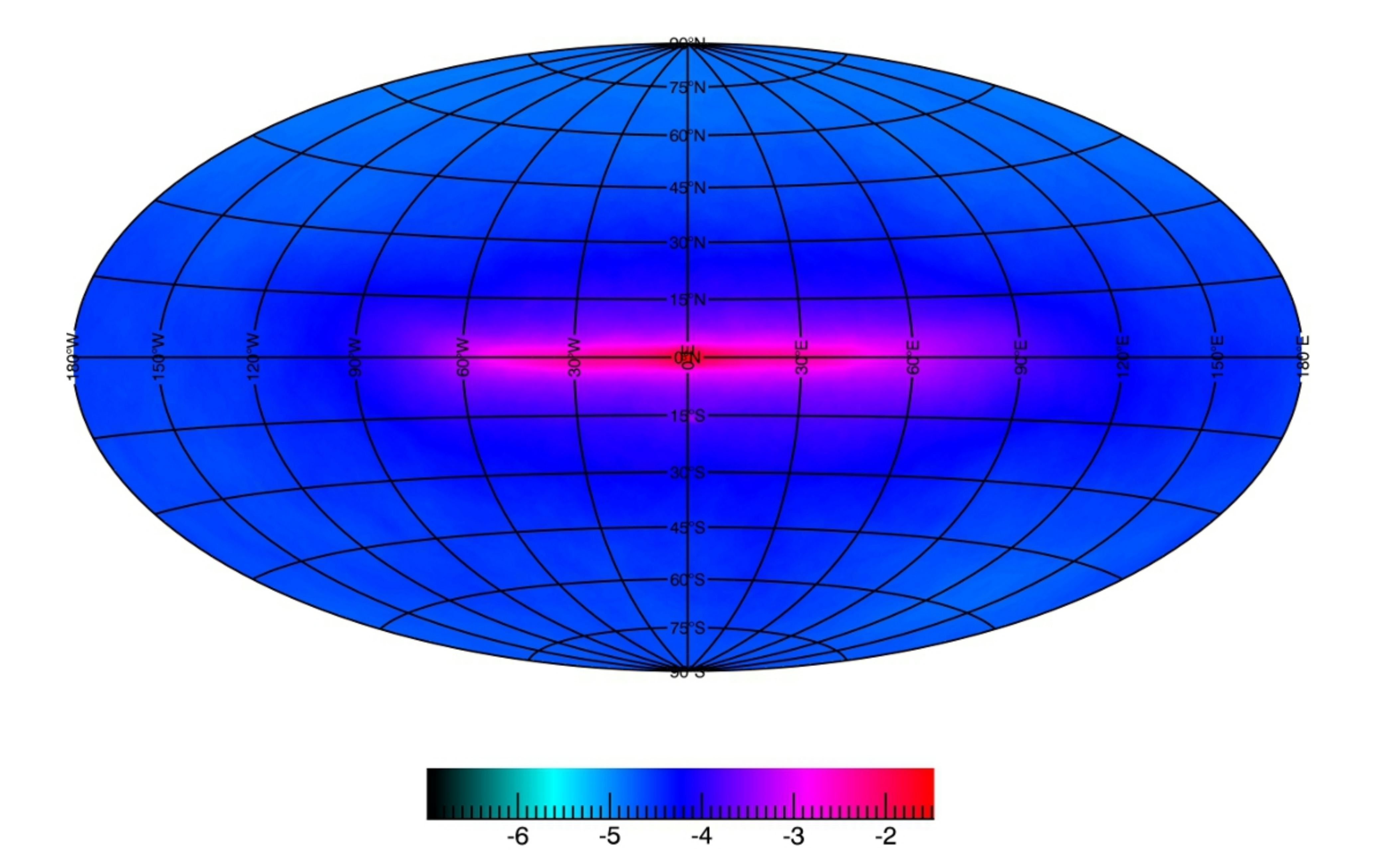}
\includegraphics[width=0.32\textwidth]{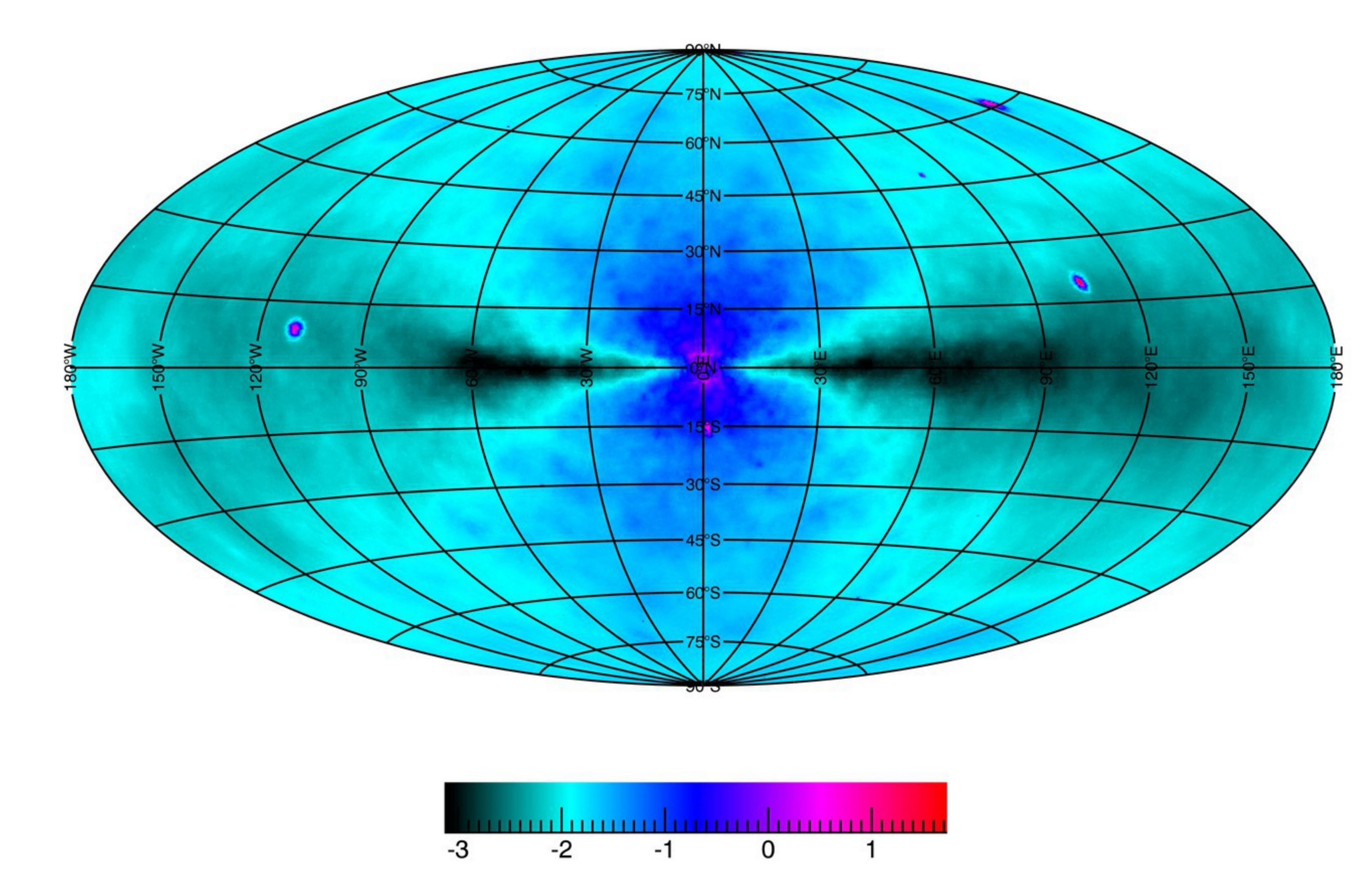}
\caption{Left: log-scaled skymap of the DM annihilation gamma-ray signal 
(${\rm log}_{10}\{ \Phi(>1\,{\rm GeV}) / {\rm cm^{-2}s^{-1}sr^{-1}}\}$), assuming 100 GeV WIMPs 
annihilating into $b\bar{b}$ quark pairs. Middle: DGRE due to CR interactions
with the interstellar gas (same units as in left panel). Right: Signal-to-background ratio
(log scale).}
\label{fig:res}
\end{figure*}

As a final series of illustrations, we have explicitly calculated the gamma-ray skymap that our 
virtual observer would measure (flux integrated above 1 GeV, smeared over an angular window of 
0.01 squared degree). In the left panel of \citefig{fig:res}, we show the contribution of DM 
annihilation, where we assumed a 100 GeV WIMP annihilating into $b\bar{b}$ quark pairs with a 
canonical thermal cross section of $\sigv = 3\times 10^{-26} \,{\rm cm^3 /s}$. This map exhibits
an almost spherical symmetry around the galactic center, a well-known property of DM halos. The 
signal concentrates in the very center, while four other hot spots are also visible. These are 
subhalos
with masses $M$ in the range $10^8$-$10^9$ \msun, and lying at distances $d$ between 20-30 kpc from 
the observer. This mass range is that of the most massive dwarf galaxy satellites of the MW. Since 
the subhalo gamma-ray flux scales like $\sim M/d^2$, we note that the spots obtained here are much 
brighter, by 2 or 3 orders of magnitude, than what we would expect from the MW satellites, which are
more distant. In the middle panel, our calculated astrophysical background is shown with the same 
level of precision, for which we see a much flatter and smoother distribution than for the 
DM-induced component. We make the ratio in the right panel: regions favorable to DM searches are
around the galactic center and a few massive subhalos, as expected.

Of course, the virtual observer would only measure
the sum of the left and middle panels, and would have difficulties in reaching such conclusions
without an accurate and reliable background model, while the background map is exact in our
case. Nevertheless, this already really allows one to assess the potential of indirect DM 
detection with gamma rays. We notably remark that while our simulation is characterized by
a significant adiabatic contraction of the DM density in the central parts of the galaxy, the 
background is still strong enough to induce a rather modest ratio. This result is rather unexpected,
since adiabatic contraction is usually thought of as a very optimistic case for 
DM signal predictions. One could still take a lighter WIMP to get a larger flux. 
Nevertheless, there is another physical reason: adiabatic contraction stems from the over-cooling 
of baryons, which also implies a large central gas density, and a high local star formation rate 
(and thereby a large local SN explosion rate and CR density). Therefore, adiabatic contraction 
amplifies 
both the DM signal and the background. Since gravitation affects DM and baryons the same, one could 
argue that the signal should still be favored because it is proportional to the squared DM density. 
Nevertheless, the background is somehow also proportional to the squared baryon density. Indeed, 
locally, the CR density is proportional to the number density of sources, which is itself 
proportional to the gas density (the star formation rate scales linearly with the gas density in our
simulation). Therefore, since the background flux is set by the product of the gas and CR densities,
this corresponds to squaring the overall baryon density (CRs are actually much more diluted than 
the gas because of diffusion, which slightly alters the argument, though keeping it qualitatively 
valid).

We have incorporated sophisticated ingredients of CR physics in a cosmological
simulation including baryons. We focused on a zoomed-in galaxy with properties
reasonably comparable to those of the MW. We have calculated the CR density self-consistently after 
having accurately localized their sources (SN-like events) in space and time, and 
convoluted it with the interstellar gas density to calculate the DGRE. We 
have determined the DGRE that a virtual Earth-like observer located in the disk would detect, 
and discussed separately the amplitude of the DM annihilation signal and that of the astrophysical 
background. We have shown that even in an adiabatically contracted halo (not necessarily to be 
compared with the MW), the signal-to-noise ratio is not necessarily large. This is due to the fact 
that both the DM and baryon densities are correlated in such a case, which induces an 
amplification in both the signal and background. Irrespective of any putative similarities
or differences with the MW, a firm conclusion is that from this unprecedented top-bottom
approach, we demonstrate that indirect DM detection is in any case challenging, whatever the
central DM shape. The best targets remain the DM subhalos, some of which are found to be observable
in our simulated galaxy, though with values of $M/d^2$ much more favorable than those of the known 
MW satellites. The central parts of the galaxy are very interesting zones to which we
will further dedicate a detailed analysis. This approach can be generalized to all cosmic tracers 
of DM annihilation, like antimatter CRs, radio and X emissions, and neutrinos. Even more globally, 
it may serve as a powerful diagnostic tool to test galaxy formation itself from its high-energy 
properties.

{\em Acknowledgments}: We warmly thank Jean-Charles Lambert for the very valuable support he provided in the computing aspects of this work.

\bibliography{mybib}

\end{document}

%% file: newcommands.tex
\newcommand{\new}[1]{{#1}}
\newcommand{\change}[1]{{#1}}

\newcommand{\citefig}[1]{Fig.~\ref{#1}}

\newcommand{\eg}{{\it e.g.}}
\newcommand{\ie}{{\it i.e.}}

\newcommand{\ben}{\begin{eqnarray}}
\newcommand{\een}{\end{eqnarray}}
\newcommand{\be}{\begin{equation}}
\newcommand{\ee}{\end{equation}}
\newcommand{\besubeq}[1]{\begin{subequations} \label{#1}}
\newcommand{\eesubeq}{\end{subequations}}

\newcommand{\msun}{\mbox{$M_{\odot}$}}

\newcommand{\sigv}{\mbox{$\langle \sigma v \rangle $}}

\newcommand{\aap}{Astron. Astroph.}

\newcommand{\apjl}{Astrophys. J. Lett.}

\newcommand{\jcap}{JCAP}
\newcommand{\mnras}{MNRAS}
 
\newcommand{\physrep}{Phys. Rept.}

%% file: nbody_cosmic_rays_nlt.bbl
\begin{thebibliography}{39}%
\makeatletter
\providecommand \@ifxundefined [1]{%
 \@ifx{#1\undefined}
}%
\providecommand \@ifnum [1]{%
 \ifnum #1\expandafter \@firstoftwo
 \else \expandafter \@secondoftwo
 \fi
}%
\providecommand \@ifx [1]{%
 \ifx #1\expandafter \@firstoftwo
 \else \expandafter \@secondoftwo
 \fi
}%
\providecommand \natexlab [1]{#1}%
\providecommand \enquote  [1]{``#1''}%
\providecommand \bibnamefont  [1]{#1}%
\providecommand \bibfnamefont [1]{#1}%
\providecommand \citenamefont [1]{#1}%
\providecommand \href@noop [0]{\@secondoftwo}%
\providecommand \href [0]{\begingroup \@sanitize@url \@href}%
\providecommand \@href[1]{\@@startlink{#1}\@@href}%
\providecommand \@@href[1]{\endgroup#1\@@endlink}%
\providecommand \@sanitize@url [0]{\catcode `\\12\catcode `\$12\catcode
  `\&12\catcode `\#12\catcode `\^12\catcode `\_12\catcode `\%12\relax}%
\providecommand \@@startlink[1]{}%
\providecommand \@@endlink[0]{}%
\providecommand \url  [0]{\begingroup\@sanitize@url \@url }%
\providecommand \@url [1]{\endgroup\@href {#1}{\urlprefix }}%
\providecommand \urlprefix  [0]{URL }%
\providecommand \Eprint [0]{\href }%
\providecommand \doibase [0]{http://dx.doi.org/}%
\providecommand \selectlanguage [0]{\@gobble}%
\providecommand \bibinfo  [0]{\@secondoftwo}%
\providecommand \bibfield  [0]{\@secondoftwo}%
\providecommand \translation [1]{[#1]}%
\providecommand \BibitemOpen [0]{}%
\providecommand \bibitemStop [0]{}%
\providecommand \bibitemNoStop [0]{.\EOS\space}%
\providecommand \EOS [0]{\spacefactor3000\relax}%
\providecommand \BibitemShut  [1]{\csname bibitem#1\endcsname}%
\let\auto@bib@innerbib\@empty
\bibitem [{\citenamefont {{Jungman}}\ \emph {et~al.}(1996)\citenamefont
  {{Jungman}}, \citenamefont {{Kamionkowski}},\ and\ \citenamefont
  {{Griest}}}]{1996PhR...267..195J}%
  \BibitemOpen
  \bibfield  {author} {\bibinfo {author} {\bibfnamefont {G.}~\bibnamefont
  {{Jungman}}}, \bibinfo {author} {\bibfnamefont {M.}~\bibnamefont
  {{Kamionkowski}}}, \ and\ \bibinfo {author} {\bibfnamefont {K.}~\bibnamefont
  {{Griest}}},\ }\href {\doibase 10.1016/0370-1573(95)00058-5} {\bibfield
  {journal} {\bibinfo  {journal} {\physrep}\ }\textbf {\bibinfo {volume}
  {267}},\ \bibinfo {pages} {195} (\bibinfo {year} {1996})},\ \Eprint
  {http://arxiv.org/abs/hep-ph/9506380} {hep-ph/9506380} \BibitemShut {NoStop}%
\bibitem [{\citenamefont {{Bergstr{\"o}m}}(2000)}]{2000RPPh...63..793B}%
  \BibitemOpen
  \bibfield  {author} {\bibinfo {author} {\bibfnamefont {L.}~\bibnamefont
  {{Bergstr{\"o}m}}},\ }\href {\doibase 10.1088/0034-4885/63/5/2r3} {\bibfield
  {journal} {\bibinfo  {journal} {Reports on Progress in Physics}\ }\textbf
  {\bibinfo {volume} {63}},\ \bibinfo {pages} {793} (\bibinfo {year} {2000})},\
  \Eprint {http://arxiv.org/abs/hep-ph/0002126} {hep-ph/0002126} \BibitemShut
  {NoStop}%
\bibitem [{\citenamefont {{Carr}}\ \emph {et~al.}(2006)\citenamefont {{Carr}},
  \citenamefont {{Lamanna}},\ and\ \citenamefont
  {{Lavalle}}}]{2006RPPh...69.2475C}%
  \BibitemOpen
  \bibfield  {author} {\bibinfo {author} {\bibfnamefont {J.}~\bibnamefont
  {{Carr}}}, \bibinfo {author} {\bibfnamefont {G.}~\bibnamefont {{Lamanna}}}, \
  and\ \bibinfo {author} {\bibfnamefont {J.}~\bibnamefont {{Lavalle}}},\ }\href
  {\doibase 10.1088/0034-4885/69/8/R05} {\bibfield  {journal} {\bibinfo
  {journal} {Reports on Progress in Physics}\ }\textbf {\bibinfo {volume}
  {69}},\ \bibinfo {pages} {2475} (\bibinfo {year} {2006})}\BibitemShut
  {NoStop}%
\bibitem [{\citenamefont {Salati}(2007)}]{Salati:2007zz}%
  \BibitemOpen
  \bibfield  {author} {\bibinfo {author} {\bibfnamefont {P.}~\bibnamefont
  {Salati}},\ }\href@noop {} {\bibfield  {journal} {\bibinfo  {journal}
  {Proceedings of Science}\ }\textbf {\bibinfo {volume} {Cargese 2007}},\
  \bibinfo {pages} {009} (\bibinfo {year} {2007})}\BibitemShut {NoStop}%
\bibitem [{\citenamefont {{Hooper}}\ \emph {et~al.}(2007)\citenamefont
  {{Hooper}}, \citenamefont {{Finkbeiner}},\ and\ \citenamefont
  {{Dobler}}}]{2007PhRvD..76h3012H}%
  \BibitemOpen
  \bibfield  {author} {\bibinfo {author} {\bibfnamefont {D.}~\bibnamefont
  {{Hooper}}}, \bibinfo {author} {\bibfnamefont {D.~P.}\ \bibnamefont
  {{Finkbeiner}}}, \ and\ \bibinfo {author} {\bibfnamefont {G.}~\bibnamefont
  {{Dobler}}},\ }\href {\doibase 10.1103/PhysRevD.76.083012} {\bibfield
  {journal} {\bibinfo  {journal} {\prd}\ }\textbf {\bibinfo {volume} {76}},\
  \bibinfo {pages} {083012} (\bibinfo {year} {2007})},\ \Eprint
  {http://arxiv.org/abs/0705.3655} {arXiv:0705.3655} \BibitemShut {NoStop}%
\bibitem [{\citenamefont {{Mertsch}}\ and\ \citenamefont
  {{Sarkar}}(2010)}]{2010JCAP...10..019M}%
  \BibitemOpen
  \bibfield  {author} {\bibinfo {author} {\bibfnamefont {P.}~\bibnamefont
  {{Mertsch}}}\ and\ \bibinfo {author} {\bibfnamefont {S.}~\bibnamefont
  {{Sarkar}}},\ }\href {\doibase 10.1088/1475-7516/2010/10/019} {\bibfield
  {journal} {\bibinfo  {journal} {\jcap}\ }\textbf {\bibinfo {volume} {10}},\
  \bibinfo {pages} {19} (\bibinfo {year} {2010})},\ \Eprint
  {http://arxiv.org/abs/1004.3056} {arXiv:1004.3056 [astro-ph.HE]} \BibitemShut
  {NoStop}%
\bibitem [{\citenamefont {{Adriani}}\ \emph {et~al.}(2009)\citenamefont
  {{Adriani}} \emph {et~al.}}]{2009Natur.458..607A}%
  \BibitemOpen
  \bibfield  {author} {\bibinfo {author} {\bibfnamefont {O.}~\bibnamefont
  {{Adriani}}} \emph {et~al.},\ }\href {\doibase 10.1038/nature07942}
  {\bibfield  {journal} {\bibinfo  {journal} {\nat}\ }\textbf {\bibinfo
  {volume} {458}},\ \bibinfo {pages} {607} (\bibinfo {year} {2009})},\ \Eprint
  {http://arxiv.org/abs/0810.4995} {arXiv:0810.4995} \BibitemShut {NoStop}%
\bibitem [{\citenamefont {{Bergstr{\"o}m}}\ \emph {et~al.}(2008)\citenamefont
  {{Bergstr{\"o}m}}, \citenamefont {{Bringmann}},\ and\ \citenamefont
  {{Edsj{\"o}}}}]{2008PhRvD..78j3520B}%
  \BibitemOpen
  \bibfield  {author} {\bibinfo {author} {\bibfnamefont {L.}~\bibnamefont
  {{Bergstr{\"o}m}}}, \bibinfo {author} {\bibfnamefont {T.}~\bibnamefont
  {{Bringmann}}}, \ and\ \bibinfo {author} {\bibfnamefont {J.}~\bibnamefont
  {{Edsj{\"o}}}},\ }\href {\doibase 10.1103/PhysRevD.78.103520} {\bibfield
  {journal} {\bibinfo  {journal} {\prd}\ }\textbf {\bibinfo {volume} {78}},\
  \bibinfo {pages} {103520} (\bibinfo {year} {2008})},\ \Eprint
  {http://arxiv.org/abs/0808.3725} {arXiv:0808.3725} \BibitemShut {NoStop}%
\bibitem [{\citenamefont {{Harding}}\ and\ \citenamefont
  {{Ramaty}}(1987)}]{1987ICRC....2...92H}%
  \BibitemOpen
  \bibfield  {author} {\bibinfo {author} {\bibfnamefont {A.~K.}\ \bibnamefont
  {{Harding}}}\ and\ \bibinfo {author} {\bibfnamefont {R.}~\bibnamefont
  {{Ramaty}}},\ }in\ \href@noop {} {\emph {\bibinfo {booktitle} {International
  Cosmic Ray Conference}}},\ \bibinfo {series} {International Cosmic Ray
  Conference}, Vol.~\bibinfo {volume} {2}\ (\bibinfo {year} {1987})\ pp.\
  \bibinfo {pages} {92--+}\BibitemShut {NoStop}%
\bibitem [{\citenamefont {{Ackermann}}\ \emph {et~al.}(2011)\citenamefont
  {{Ackermann}} \emph {et~al.}}]{2011PhRvL.107x1302A}%
  \BibitemOpen
  \bibfield  {author} {\bibinfo {author} {\bibfnamefont {M.}~\bibnamefont
  {{Ackermann}}} \emph {et~al.},\ }\href {\doibase
  10.1103/PhysRevLett.107.241302} {\bibfield  {journal} {\bibinfo  {journal}
  {Physical Review Letters}\ }\textbf {\bibinfo {volume} {107}},\ \bibinfo
  {eid} {241302} (\bibinfo {year} {2011})},\ \Eprint
  {http://arxiv.org/abs/1108.3546} {arXiv:1108.3546 [astro-ph.HE]} \BibitemShut
  {NoStop}%
\bibitem [{\citenamefont {{Ackermann}}\ \emph {et~al.}(2012)\citenamefont
  {{Ackermann}} \emph {et~al.}}]{2012ApJ...747..121A}%
  \BibitemOpen
  \bibfield  {author} {\bibinfo {author} {\bibfnamefont {M.}~\bibnamefont
  {{Ackermann}}} \emph {et~al.},\ }\href {\doibase 10.1088/0004-637X/747/2/121}
  {\bibfield  {journal} {\bibinfo  {journal} {\apj}\ }\textbf {\bibinfo
  {volume} {747}},\ \bibinfo {eid} {121} (\bibinfo {year} {2012})},\ \Eprint
  {http://arxiv.org/abs/1201.2691} {arXiv:1201.2691 [astro-ph.HE]} \BibitemShut
  {NoStop}%
\bibitem [{\citenamefont {{Bergstr{\"o}m}}\ \emph {et~al.}(1998)\citenamefont
  {{Bergstr{\"o}m}}, \citenamefont {{Ullio}},\ and\ \citenamefont
  {{Buckley}}}]{1998APh.....9..137B}%
  \BibitemOpen
  \bibfield  {author} {\bibinfo {author} {\bibfnamefont {L.}~\bibnamefont
  {{Bergstr{\"o}m}}}, \bibinfo {author} {\bibfnamefont {P.}~\bibnamefont
  {{Ullio}}}, \ and\ \bibinfo {author} {\bibfnamefont {J.~H.}\ \bibnamefont
  {{Buckley}}},\ }\href@noop {} {\bibfield  {journal} {\bibinfo  {journal}
  {Astroparticle Physics}\ }\textbf {\bibinfo {volume} {9}},\ \bibinfo {pages}
  {137} (\bibinfo {year} {1998})},\ \Eprint
  {http://arxiv.org/abs/astro-ph/9712318} {astro-ph/9712318} \BibitemShut
  {NoStop}%
\bibitem [{\citenamefont {{Pieri}}\ \emph {et~al.}(2011)\citenamefont
  {{Pieri}}, \citenamefont {{Lavalle}}, \citenamefont {{Bertone}},\ and\
  \citenamefont {{Branchini}}}]{2011PhRvD..83b3518P}%
  \BibitemOpen
  \bibfield  {author} {\bibinfo {author} {\bibfnamefont {L.}~\bibnamefont
  {{Pieri}}}, \bibinfo {author} {\bibfnamefont {J.}~\bibnamefont {{Lavalle}}},
  \bibinfo {author} {\bibfnamefont {G.}~\bibnamefont {{Bertone}}}, \ and\
  \bibinfo {author} {\bibfnamefont {E.}~\bibnamefont {{Branchini}}},\ }\href
  {\doibase 10.1103/PhysRevD.83.023518} {\bibfield  {journal} {\bibinfo
  {journal} {\prd}\ }\textbf {\bibinfo {volume} {83}},\ \bibinfo {pages}
  {023518} (\bibinfo {year} {2011})},\ \Eprint {http://arxiv.org/abs/0908.0195}
  {arXiv:0908.0195 [astro-ph.HE]} \BibitemShut {NoStop}%
\bibitem [{\citenamefont {{The Fermi-LAT
  Collaboration}}(2012)}]{2012arXiv1202.4039T}%
  \BibitemOpen
  \bibfield  {author} {\bibinfo {author} {\bibnamefont {{The Fermi-LAT
  Collaboration}}},\ }\href@noop {} {\bibfield  {journal} {\bibinfo  {journal}
  {ArXiv e-prints}\ } (\bibinfo {year} {2012})},\ \Eprint
  {http://arxiv.org/abs/1202.4039} {arXiv:1202.4039 [astro-ph.HE]} \BibitemShut
  {NoStop}%
\bibitem [{\citenamefont {{Delahaye}}\ \emph {et~al.}(2011)\citenamefont
  {{Delahaye}}, \citenamefont {{Fiasson}}, \citenamefont {{Pohl}},\ and\
  \citenamefont {{Salati}}}]{2011A&A...531A..37D}%
  \BibitemOpen
  \bibfield  {author} {\bibinfo {author} {\bibfnamefont {T.}~\bibnamefont
  {{Delahaye}}}, \bibinfo {author} {\bibfnamefont {A.}~\bibnamefont
  {{Fiasson}}}, \bibinfo {author} {\bibfnamefont {M.}~\bibnamefont {{Pohl}}}, \
  and\ \bibinfo {author} {\bibfnamefont {P.}~\bibnamefont {{Salati}}},\ }\href
  {\doibase 10.1051/0004-6361/201116647} {\bibfield  {journal} {\bibinfo
  {journal} {\aap}\ }\textbf {\bibinfo {volume} {531}},\ \bibinfo {eid} {A37}
  (\bibinfo {year} {2011})},\ \Eprint {http://arxiv.org/abs/1102.0744}
  {arXiv:1102.0744 [astro-ph.HE]} \BibitemShut {NoStop}%
\bibitem [{\citenamefont {{Beacom}}\ \emph {et~al.}(2005)\citenamefont
  {{Beacom}}, \citenamefont {{Bell}},\ and\ \citenamefont
  {{Bertone}}}]{2005PhRvL..94q1301B}%
  \BibitemOpen
  \bibfield  {author} {\bibinfo {author} {\bibfnamefont {J.~F.}\ \bibnamefont
  {{Beacom}}}, \bibinfo {author} {\bibfnamefont {N.~F.}\ \bibnamefont
  {{Bell}}}, \ and\ \bibinfo {author} {\bibfnamefont {G.}~\bibnamefont
  {{Bertone}}},\ }\href {\doibase 10.1103/PhysRevLett.94.171301} {\bibfield
  {journal} {\bibinfo  {journal} {Physical Review Letters}\ }\textbf {\bibinfo
  {volume} {94}},\ \bibinfo {pages} {171301} (\bibinfo {year} {2005})},\
  \Eprint {http://arxiv.org/abs/astro-ph/0409403} {astro-ph/0409403}
  \BibitemShut {NoStop}%
\bibitem [{\citenamefont {{Bringmann}}\ \emph {et~al.}(2012)\citenamefont
  {{Bringmann}}, \citenamefont {{Huang}}, \citenamefont {{Ibarra}},
  \citenamefont {{Vogl}},\ and\ \citenamefont
  {{Weniger}}}]{2012arXiv1203.1312B}%
  \BibitemOpen
  \bibfield  {author} {\bibinfo {author} {\bibfnamefont {T.}~\bibnamefont
  {{Bringmann}}}, \bibinfo {author} {\bibfnamefont {X.}~\bibnamefont
  {{Huang}}}, \bibinfo {author} {\bibfnamefont {A.}~\bibnamefont {{Ibarra}}},
  \bibinfo {author} {\bibfnamefont {S.}~\bibnamefont {{Vogl}}}, \ and\ \bibinfo
  {author} {\bibfnamefont {C.}~\bibnamefont {{Weniger}}},\ }\href@noop {}
  {\bibfield  {journal} {\bibinfo  {journal} {ArXiv e-prints}\ } (\bibinfo
  {year} {2012})},\ \Eprint {http://arxiv.org/abs/1203.1312} {arXiv:1203.1312
  [hep-ph]} \BibitemShut {NoStop}%
\bibitem [{\citenamefont {{Pohl}}\ \emph {et~al.}(2008)\citenamefont {{Pohl}},
  \citenamefont {{Englmaier}},\ and\ \citenamefont
  {{Bissantz}}}]{2008ApJ...677..283P}%
  \BibitemOpen
  \bibfield  {author} {\bibinfo {author} {\bibfnamefont {M.}~\bibnamefont
  {{Pohl}}}, \bibinfo {author} {\bibfnamefont {P.}~\bibnamefont {{Englmaier}}},
  \ and\ \bibinfo {author} {\bibfnamefont {N.}~\bibnamefont {{Bissantz}}},\
  }\href {\doibase 10.1086/529004} {\bibfield  {journal} {\bibinfo  {journal}
  {\apj}\ }\textbf {\bibinfo {volume} {677}},\ \bibinfo {pages} {283} (\bibinfo
  {year} {2008})},\ \Eprint {http://arxiv.org/abs/0712.4264} {arXiv:0712.4264}
  \BibitemShut {NoStop}%
\bibitem [{\citenamefont {{Blumenthal}}\ \emph {et~al.}(1986)\citenamefont
  {{Blumenthal}}, \citenamefont {{Faber}}, \citenamefont {{Flores}},\ and\
  \citenamefont {{Primack}}}]{1986ApJ...301...27B}%
  \BibitemOpen
  \bibfield  {author} {\bibinfo {author} {\bibfnamefont {G.~R.}\ \bibnamefont
  {{Blumenthal}}}, \bibinfo {author} {\bibfnamefont {S.~M.}\ \bibnamefont
  {{Faber}}}, \bibinfo {author} {\bibfnamefont {R.}~\bibnamefont {{Flores}}}, \
  and\ \bibinfo {author} {\bibfnamefont {J.~R.}\ \bibnamefont {{Primack}}},\
  }\href {\doibase 10.1086/163867} {\bibfield  {journal} {\bibinfo  {journal}
  {\apj}\ }\textbf {\bibinfo {volume} {301}},\ \bibinfo {pages} {27} (\bibinfo
  {year} {1986})}\BibitemShut {NoStop}%
\bibitem [{\citenamefont {{Gnedin}}\ \emph {et~al.}(2004)\citenamefont
  {{Gnedin}}, \citenamefont {{Kravtsov}}, \citenamefont {{Klypin}},\ and\
  \citenamefont {{Nagai}}}]{2004ApJ...616...16G}%
  \BibitemOpen
  \bibfield  {author} {\bibinfo {author} {\bibfnamefont {O.~Y.}\ \bibnamefont
  {{Gnedin}}}, \bibinfo {author} {\bibfnamefont {A.~V.}\ \bibnamefont
  {{Kravtsov}}}, \bibinfo {author} {\bibfnamefont {A.~A.}\ \bibnamefont
  {{Klypin}}}, \ and\ \bibinfo {author} {\bibfnamefont {D.}~\bibnamefont
  {{Nagai}}},\ }\href {\doibase 10.1086/424914} {\bibfield  {journal} {\bibinfo
   {journal} {\apj}\ }\textbf {\bibinfo {volume} {616}},\ \bibinfo {pages} {16}
  (\bibinfo {year} {2004})},\ \Eprint {http://arxiv.org/abs/astro-ph/0406247}
  {astro-ph/0406247} \BibitemShut {NoStop}%
\bibitem [{\citenamefont {{Teyssier}}(2002)}]{2002A&A...385..337T}%
  \BibitemOpen
  \bibfield  {author} {\bibinfo {author} {\bibfnamefont {R.}~\bibnamefont
  {{Teyssier}}},\ }\href {\doibase 10.1051/0004-6361:20011817} {\bibfield
  {journal} {\bibinfo  {journal} {\aap}\ }\textbf {\bibinfo {volume} {385}},\
  \bibinfo {pages} {337} (\bibinfo {year} {2002})},\ \Eprint
  {http://arxiv.org/abs/astro-ph/0111367} {astro-ph/0111367} \BibitemShut
  {NoStop}%
\bibitem [{\citenamefont {{Ling}}\ \emph {et~al.}(2010)\citenamefont {{Ling}},
  \citenamefont {{Nezri}}, \citenamefont {{Athanassoula}},\ and\ \citenamefont
  {{Teyssier}}}]{2010JCAP...02..012L}%
  \BibitemOpen
  \bibfield  {author} {\bibinfo {author} {\bibfnamefont {F.-S.}\ \bibnamefont
  {{Ling}}}, \bibinfo {author} {\bibfnamefont {E.}~\bibnamefont {{Nezri}}},
  \bibinfo {author} {\bibfnamefont {E.}~\bibnamefont {{Athanassoula}}}, \ and\
  \bibinfo {author} {\bibfnamefont {R.}~\bibnamefont {{Teyssier}}},\ }\href
  {\doibase 10.1088/1475-7516/2010/02/012} {\bibfield  {journal} {\bibinfo
  {journal} {\jcap}\ }\textbf {\bibinfo {volume} {2}},\ \bibinfo {pages} {12}
  (\bibinfo {year} {2010})},\ \Eprint {http://arxiv.org/abs/0909.2028}
  {arXiv:0909.2028 [astro-ph.GA]} \BibitemShut {NoStop}%
\bibitem [{\citenamefont {{Diemand}}\ \emph {et~al.}(2008)\citenamefont
  {{Diemand}}, \citenamefont {{Kuhlen}}, \citenamefont {{Madau}}, \citenamefont
  {{Zemp}}, \citenamefont {{Moore}}, \citenamefont {{Potter}},\ and\
  \citenamefont {{Stadel}}}]{2008Natur.454..735D}%
  \BibitemOpen
  \bibfield  {author} {\bibinfo {author} {\bibfnamefont {J.}~\bibnamefont
  {{Diemand}}}, \bibinfo {author} {\bibfnamefont {M.}~\bibnamefont {{Kuhlen}}},
  \bibinfo {author} {\bibfnamefont {P.}~\bibnamefont {{Madau}}}, \bibinfo
  {author} {\bibfnamefont {M.}~\bibnamefont {{Zemp}}}, \bibinfo {author}
  {\bibfnamefont {B.}~\bibnamefont {{Moore}}}, \bibinfo {author} {\bibfnamefont
  {D.}~\bibnamefont {{Potter}}}, \ and\ \bibinfo {author} {\bibfnamefont
  {J.}~\bibnamefont {{Stadel}}},\ }\href {\doibase 10.1038/nature07153}
  {\bibfield  {journal} {\bibinfo  {journal} {\nat}\ }\textbf {\bibinfo
  {volume} {454}},\ \bibinfo {pages} {735} (\bibinfo {year} {2008})},\ \Eprint
  {http://arxiv.org/abs/0805.1244} {arXiv:0805.1244} \BibitemShut {NoStop}%
\bibitem [{\citenamefont {{Springel}}\ \emph {et~al.}(2008)\citenamefont
  {{Springel}}, \citenamefont {{White}}, \citenamefont {{Frenk}}, \citenamefont
  {{Navarro}}, \citenamefont {{Jenkins}}, \citenamefont {{Vogelsberger}},
  \citenamefont {{Wang}}, \citenamefont {{Ludlow}},\ and\ \citenamefont
  {{Helmi}}}]{2008Natur.456...73S}%
  \BibitemOpen
  \bibfield  {author} {\bibinfo {author} {\bibfnamefont {V.}~\bibnamefont
  {{Springel}}}, \bibinfo {author} {\bibfnamefont {S.~D.~M.}\ \bibnamefont
  {{White}}}, \bibinfo {author} {\bibfnamefont {C.~S.}\ \bibnamefont
  {{Frenk}}}, \bibinfo {author} {\bibfnamefont {J.~F.}\ \bibnamefont
  {{Navarro}}}, \bibinfo {author} {\bibfnamefont {A.}~\bibnamefont
  {{Jenkins}}}, \bibinfo {author} {\bibfnamefont {M.}~\bibnamefont
  {{Vogelsberger}}}, \bibinfo {author} {\bibfnamefont {J.}~\bibnamefont
  {{Wang}}}, \bibinfo {author} {\bibfnamefont {A.}~\bibnamefont {{Ludlow}}}, \
  and\ \bibinfo {author} {\bibfnamefont {A.}~\bibnamefont {{Helmi}}},\ }\href
  {\doibase 10.1038/nature07411} {\bibfield  {journal} {\bibinfo  {journal}
  {\nat}\ }\textbf {\bibinfo {volume} {456}},\ \bibinfo {pages} {73} (\bibinfo
  {year} {2008})},\ \Eprint {http://arxiv.org/abs/0809.0894} {arXiv:0809.0894}
  \BibitemShut {NoStop}%
\bibitem [{\citenamefont {{Lavalle}}\ \emph {et~al.}(2008)\citenamefont
  {{Lavalle}}, \citenamefont {{Yuan}}, \citenamefont {{Maurin}},\ and\
  \citenamefont {{Bi}}}]{2008A&A...479..427L}%
  \BibitemOpen
  \bibfield  {author} {\bibinfo {author} {\bibfnamefont {J.}~\bibnamefont
  {{Lavalle}}}, \bibinfo {author} {\bibfnamefont {Q.}~\bibnamefont {{Yuan}}},
  \bibinfo {author} {\bibfnamefont {D.}~\bibnamefont {{Maurin}}}, \ and\
  \bibinfo {author} {\bibfnamefont {X.-J.}\ \bibnamefont {{Bi}}},\ }\href
  {\doibase 10.1051/0004-6361:20078723} {\bibfield  {journal} {\bibinfo
  {journal} {\aap}\ }\textbf {\bibinfo {volume} {479}},\ \bibinfo {pages} {427}
  (\bibinfo {year} {2008})},\ \Eprint {http://arxiv.org/abs/0709.3634}
  {0709.3634} \BibitemShut {NoStop}%
\bibitem [{\citenamefont {{Governato}}\ \emph {et~al.}(2010)\citenamefont
  {{Governato}}, \citenamefont {{Brook}}, \citenamefont {{Mayer}},
  \citenamefont {{Brooks}}, \citenamefont {{Rhee}}, \citenamefont {{Wadsley}},
  \citenamefont {{Jonsson}}, \citenamefont {{Willman}}, \citenamefont
  {{Stinson}}, \citenamefont {{Quinn}},\ and\ \citenamefont
  {{Madau}}}]{2010Natur.463..203G}%
  \BibitemOpen
  \bibfield  {author} {\bibinfo {author} {\bibfnamefont {F.}~\bibnamefont
  {{Governato}}}, \bibinfo {author} {\bibfnamefont {C.}~\bibnamefont
  {{Brook}}}, \bibinfo {author} {\bibfnamefont {L.}~\bibnamefont {{Mayer}}},
  \bibinfo {author} {\bibfnamefont {A.}~\bibnamefont {{Brooks}}}, \bibinfo
  {author} {\bibfnamefont {G.}~\bibnamefont {{Rhee}}}, \bibinfo {author}
  {\bibfnamefont {J.}~\bibnamefont {{Wadsley}}}, \bibinfo {author}
  {\bibfnamefont {P.}~\bibnamefont {{Jonsson}}}, \bibinfo {author}
  {\bibfnamefont {B.}~\bibnamefont {{Willman}}}, \bibinfo {author}
  {\bibfnamefont {G.}~\bibnamefont {{Stinson}}}, \bibinfo {author}
  {\bibfnamefont {T.}~\bibnamefont {{Quinn}}}, \ and\ \bibinfo {author}
  {\bibfnamefont {P.}~\bibnamefont {{Madau}}},\ }\href {\doibase
  10.1038/nature08640} {\bibfield  {journal} {\bibinfo  {journal} {\nat}\
  }\textbf {\bibinfo {volume} {463}},\ \bibinfo {pages} {203} (\bibinfo {year}
  {2010})}\BibitemShut {NoStop}%
\bibitem [{\citenamefont {{Scannapieco}}\ \emph {et~al.}(2011)\citenamefont
  {{Scannapieco}} \emph {et~al.}}]{2011arXiv1112.0315S}%
  \BibitemOpen
  \bibfield  {author} {\bibinfo {author} {\bibfnamefont {C.}~\bibnamefont
  {{Scannapieco}}} \emph {et~al.},\ }\href@noop {} {\bibfield  {journal}
  {\bibinfo  {journal} {ArXiv e-prints}\ } (\bibinfo {year} {2011})},\ \Eprint
  {http://arxiv.org/abs/1112.0315} {arXiv:1112.0315 [astro-ph.GA]} \BibitemShut
  {NoStop}%
\bibitem [{\citenamefont {{Uhlig}}\ \emph {et~al.}(2012)\citenamefont
  {{Uhlig}}, \citenamefont {{Pfrommer}}, \citenamefont {{Sharma}},
  \citenamefont {{Nath}}, \citenamefont {{Ensslin}},\ and\ \citenamefont
  {{Springel}}}]{2012arXiv1203.1038U}%
  \BibitemOpen
  \bibfield  {author} {\bibinfo {author} {\bibfnamefont {M.}~\bibnamefont
  {{Uhlig}}}, \bibinfo {author} {\bibfnamefont {C.}~\bibnamefont {{Pfrommer}}},
  \bibinfo {author} {\bibfnamefont {M.}~\bibnamefont {{Sharma}}}, \bibinfo
  {author} {\bibfnamefont {B.~B.}\ \bibnamefont {{Nath}}}, \bibinfo {author}
  {\bibfnamefont {T.~A.}\ \bibnamefont {{Ensslin}}}, \ and\ \bibinfo {author}
  {\bibfnamefont {V.}~\bibnamefont {{Springel}}},\ }\href@noop {} {\bibfield
  {journal} {\bibinfo  {journal} {ArXiv e-prints}\ } (\bibinfo {year}
  {2012})},\ \Eprint {http://arxiv.org/abs/1203.1038} {arXiv:1203.1038
  [astro-ph.CO]} \BibitemShut {NoStop}%
\bibitem [{\citenamefont {{Berezinskii}}\ \emph {et~al.}(1990)\citenamefont
  {{Berezinskii}}, \citenamefont {{Bulanov}}, \citenamefont {{Dogiel}},\ and\
  \citenamefont {{Ptuskin}}}]{berezinsky_book_90}%
  \BibitemOpen
  \bibfield  {author} {\bibinfo {author} {\bibfnamefont {V.~S.}\ \bibnamefont
  {{Berezinskii}}}, \bibinfo {author} {\bibfnamefont {S.~V.}\ \bibnamefont
  {{Bulanov}}}, \bibinfo {author} {\bibfnamefont {V.~A.}\ \bibnamefont
  {{Dogiel}}}, \ and\ \bibinfo {author} {\bibfnamefont {V.~S.}\ \bibnamefont
  {{Ptuskin}}},\ }\href@noop {} {\emph {\bibinfo {title} {{Astrophysics of
  cosmic rays}}}}\ (\bibinfo  {publisher} {Amsterdam: North-Holland, edited by
  Ginzburg, V.L.},\ \bibinfo {year} {1990})\BibitemShut {NoStop}%
\bibitem [{\citenamefont {{Hanasz}}\ \emph {et~al.}(2004)\citenamefont
  {{Hanasz}}, \citenamefont {{Kowal}}, \citenamefont {{Otmianowska-Mazur}},\
  and\ \citenamefont {{Lesch}}}]{2004ApJ...605L..33H}%
  \BibitemOpen
  \bibfield  {author} {\bibinfo {author} {\bibfnamefont {M.}~\bibnamefont
  {{Hanasz}}}, \bibinfo {author} {\bibfnamefont {G.}~\bibnamefont {{Kowal}}},
  \bibinfo {author} {\bibfnamefont {K.}~\bibnamefont {{Otmianowska-Mazur}}}, \
  and\ \bibinfo {author} {\bibfnamefont {H.}~\bibnamefont {{Lesch}}},\ }\href
  {\doibase 10.1086/420697} {\bibfield  {journal} {\bibinfo  {journal} {\apjl}\
  }\textbf {\bibinfo {volume} {605}},\ \bibinfo {pages} {L33} (\bibinfo {year}
  {2004})},\ \Eprint {http://arxiv.org/abs/arXiv:astro-ph/0402662}
  {arXiv:astro-ph/0402662} \BibitemShut {NoStop}%
\bibitem [{\citenamefont {{Dubois}}\ and\ \citenamefont
  {{Teyssier}}(2008)}]{2008A&A...482L..13D}%
  \BibitemOpen
  \bibfield  {author} {\bibinfo {author} {\bibfnamefont {Y.}~\bibnamefont
  {{Dubois}}}\ and\ \bibinfo {author} {\bibfnamefont {R.}~\bibnamefont
  {{Teyssier}}},\ }\href {\doibase 10.1051/0004-6361:200809513} {\bibfield
  {journal} {\bibinfo  {journal} {\aap}\ }\textbf {\bibinfo {volume} {482}},\
  \bibinfo {pages} {L13} (\bibinfo {year} {2008})},\ \Eprint
  {http://arxiv.org/abs/0802.0490} {arXiv:0802.0490} \BibitemShut {NoStop}%
\bibitem [{\citenamefont {{Dubois}}\ and\ \citenamefont
  {{Teyssier}}(2010)}]{2010A&A...523A..72D}%
  \BibitemOpen
  \bibfield  {author} {\bibinfo {author} {\bibfnamefont {Y.}~\bibnamefont
  {{Dubois}}}\ and\ \bibinfo {author} {\bibfnamefont {R.}~\bibnamefont
  {{Teyssier}}},\ }\href {\doibase 10.1051/0004-6361/200913014} {\bibfield
  {journal} {\bibinfo  {journal} {\aap}\ }\textbf {\bibinfo {volume} {523}},\
  \bibinfo {eid} {A72} (\bibinfo {year} {2010})},\ \Eprint
  {http://arxiv.org/abs/0908.3862} {arXiv:0908.3862 [astro-ph.CO]} \BibitemShut
  {NoStop}%
\bibitem [{\citenamefont {{Beck}}\ \emph {et~al.}(2012)\citenamefont {{Beck}},
  \citenamefont {{Lesch}}, \citenamefont {{Dolag}}, \citenamefont {{Kotarba}},
  \citenamefont {{Geng}},\ and\ \citenamefont
  {{Stasyszyn}}}]{2012MNRAS.422.2152B}%
  \BibitemOpen
  \bibfield  {author} {\bibinfo {author} {\bibfnamefont {A.~M.}\ \bibnamefont
  {{Beck}}}, \bibinfo {author} {\bibfnamefont {H.}~\bibnamefont {{Lesch}}},
  \bibinfo {author} {\bibfnamefont {K.}~\bibnamefont {{Dolag}}}, \bibinfo
  {author} {\bibfnamefont {H.}~\bibnamefont {{Kotarba}}}, \bibinfo {author}
  {\bibfnamefont {A.}~\bibnamefont {{Geng}}}, \ and\ \bibinfo {author}
  {\bibfnamefont {F.~A.}\ \bibnamefont {{Stasyszyn}}},\ }\href {\doibase
  10.1111/j.1365-2966.2012.20759.x} {\bibfield  {journal} {\bibinfo  {journal}
  {\mnras}\ }\textbf {\bibinfo {volume} {422}},\ \bibinfo {pages} {2152}
  (\bibinfo {year} {2012})},\ \Eprint {http://arxiv.org/abs/1202.3349}
  {arXiv:1202.3349 [astro-ph.CO]} \BibitemShut {NoStop}%
\bibitem [{\citenamefont {{Strong}}\ and\ \citenamefont
  {{Moskalenko}}(1998)}]{1998ApJ...509..212S}%
  \BibitemOpen
  \bibfield  {author} {\bibinfo {author} {\bibfnamefont {A.~W.}\ \bibnamefont
  {{Strong}}}\ and\ \bibinfo {author} {\bibfnamefont {I.~V.}\ \bibnamefont
  {{Moskalenko}}},\ }\href {\doibase 10.1086/306470} {\bibfield  {journal}
  {\bibinfo  {journal} {\apj}\ }\textbf {\bibinfo {volume} {509}},\ \bibinfo
  {pages} {212} (\bibinfo {year} {1998})},\ \Eprint
  {http://arxiv.org/abs/astro-ph/9807150} {astro-ph/9807150} \BibitemShut
  {NoStop}%
\bibitem [{\citenamefont {{Maurin}}\ \emph {et~al.}(2001)\citenamefont
  {{Maurin}}, \citenamefont {{Donato}}, \citenamefont {{Taillet}},\ and\
  \citenamefont {{Salati}}}]{2001ApJ...555..585M}%
  \BibitemOpen
  \bibfield  {author} {\bibinfo {author} {\bibfnamefont {D.}~\bibnamefont
  {{Maurin}}}, \bibinfo {author} {\bibfnamefont {F.}~\bibnamefont {{Donato}}},
  \bibinfo {author} {\bibfnamefont {R.}~\bibnamefont {{Taillet}}}, \ and\
  \bibinfo {author} {\bibfnamefont {P.}~\bibnamefont {{Salati}}},\ }\href
  {\doibase 10.1086/321496} {\bibfield  {journal} {\bibinfo  {journal} {\apj}\
  }\textbf {\bibinfo {volume} {555}},\ \bibinfo {pages} {585} (\bibinfo {year}
  {2001})},\ \Eprint {http://arxiv.org/abs/astro-ph/0101231} {astro-ph/0101231}
  \BibitemShut {NoStop}%
\bibitem [{\citenamefont {{Evoli}}\ \emph {et~al.}(2008)\citenamefont
  {{Evoli}}, \citenamefont {{Gaggero}}, \citenamefont {{Grasso}},\ and\
  \citenamefont {{Maccione}}}]{2008JCAP...10..018E}%
  \BibitemOpen
  \bibfield  {author} {\bibinfo {author} {\bibfnamefont {C.}~\bibnamefont
  {{Evoli}}}, \bibinfo {author} {\bibfnamefont {D.}~\bibnamefont {{Gaggero}}},
  \bibinfo {author} {\bibfnamefont {D.}~\bibnamefont {{Grasso}}}, \ and\
  \bibinfo {author} {\bibfnamefont {L.}~\bibnamefont {{Maccione}}},\ }\href
  {\doibase 10.1088/1475-7516/2008/10/018} {\bibfield  {journal} {\bibinfo
  {journal} {Journal of Cosmology and Astro-Particle Physics}\ }\textbf
  {\bibinfo {volume} {10}},\ \bibinfo {pages} {18} (\bibinfo {year} {2008})},\
  \Eprint {http://arxiv.org/abs/0807.4730} {arXiv:0807.4730} \BibitemShut
  {NoStop}%
\bibitem [{\citenamefont {{Kamae}}\ \emph {et~al.}(2006)\citenamefont
  {{Kamae}}, \citenamefont {{Karlsson}}, \citenamefont {{Mizuno}},
  \citenamefont {{Abe}},\ and\ \citenamefont {{Koi}}}]{2006ApJ...647..692K}%
  \BibitemOpen
  \bibfield  {author} {\bibinfo {author} {\bibfnamefont {T.}~\bibnamefont
  {{Kamae}}}, \bibinfo {author} {\bibfnamefont {N.}~\bibnamefont {{Karlsson}}},
  \bibinfo {author} {\bibfnamefont {T.}~\bibnamefont {{Mizuno}}}, \bibinfo
  {author} {\bibfnamefont {T.}~\bibnamefont {{Abe}}}, \ and\ \bibinfo {author}
  {\bibfnamefont {T.}~\bibnamefont {{Koi}}},\ }\href {\doibase 10.1086/505189}
  {\bibfield  {journal} {\bibinfo  {journal} {\apj}\ }\textbf {\bibinfo
  {volume} {647}},\ \bibinfo {pages} {692} (\bibinfo {year} {2006})},\ \Eprint
  {http://arxiv.org/abs/astro-ph/0605581} {astro-ph/0605581} \BibitemShut
  {NoStop}%
\bibitem [{\citenamefont {{Norbury}}\ and\ \citenamefont
  {{Townsend}}(2007)}]{2007NIMPB.254..187N}%
  \BibitemOpen
  \bibfield  {author} {\bibinfo {author} {\bibfnamefont {J.~W.}\ \bibnamefont
  {{Norbury}}}\ and\ \bibinfo {author} {\bibfnamefont {L.~W.}\ \bibnamefont
  {{Townsend}}},\ }\href {\doibase 10.1016/j.nimb.2006.11.054} {\bibfield
  {journal} {\bibinfo  {journal} {Nuclear Instruments and Methods in Physics
  Research B}\ }\textbf {\bibinfo {volume} {254}},\ \bibinfo {pages} {187}
  (\bibinfo {year} {2007})},\ \Eprint {http://arxiv.org/abs/nucl-th/0612081}
  {nucl-th/0612081} \BibitemShut {NoStop}%
\bibitem [{\citenamefont {{Ferri{\`e}re}}(2001)}]{2001RvMP...73.1031F}%
  \BibitemOpen
  \bibfield  {author} {\bibinfo {author} {\bibfnamefont {K.~M.}\ \bibnamefont
  {{Ferri{\`e}re}}},\ }\href {\doibase 10.1103/RevModPhys.73.1031} {\bibfield
  {journal} {\bibinfo  {journal} {Reviews of Modern Physics}\ }\textbf
  {\bibinfo {volume} {73}},\ \bibinfo {pages} {1031} (\bibinfo {year}
  {2001})},\ \Eprint {http://arxiv.org/abs/astro-ph/0106359} {astro-ph/0106359}
  \BibitemShut {NoStop}%
\end{thebibliography}%
